\def\full{1} 
\ifnum\full=0
\documentclass{llncs}
\ifnum\full=1
\usepackage{amsmath}
\fi
\begin{document}
\mainmatter              
\title{Robust Locally Testable Codes and Products of Codes}
\titlerunning{Robust LTCs}  
%
\author{Eli Ben-Sasson\inst{1} \and Madhu Sudan\inst{2}}
\authorrunning{Ivar Ekeland et al.}   
%
\tocauthor{Eli Ben-Sasson (Radcliffe IAS) and
Madhu Sudan (MIT and Radcliffe IAS)}
\institute{Radcliffe Institute for Advanced Study,
34 Concord Avenue, Cambridge, MA 02138, USA.
\email{eli@eecs.harvard.edu}.
\and
MIT and Radcliffe IAS, The Stata Center Rm. G640, 32 Vassar Street, Cambridge, MA 02139, USA.
\email{madhu@mit.edu}.}

\maketitle              

\else
\documentclass[11pt]{article}
\usepackage{fullpage}
\usepackage{amsfonts}
\usepackage{amsmath}
\usepackage{amssymb}

\hbadness=10000
\vbadness=10000
\parskip=\medskipamount
\parindent=0in

\newtheorem{theorem}{Theorem}[section]
\newtheorem{lemma}[theorem]{Lemma}
\newtheorem{proposition}[theorem]{Proposition}
\newtheorem{definition}[theorem]{Definition}

\newcount\proofqeded
\newcount\proofended
\def\qed{ \mbox{\ \vrule width0.6ex height1em depth0cm}
\global\advance\proofqeded by 1 }
\newenvironment{proof}%
 {\proofstart}%
  {\ifnum\proofqeded=\proofended\qed\fi \global\advance\proofended by 1
  \medskip}
\makeatletter
\def\proofstart{\@ifnextchar[{\@oprf}{\@nprf}}
\def\@oprf[#1]{\paragraph{Proof of #1:}}
\def\@nprf{\paragraph{Proof:}}
\makeatother

\begin{document}

\title{Robust Locally Testable Codes and Products of Codes}
\author{%
Eli Ben-Sasson%
\thanks{Radcliffe Institute of Advanced Study,
34 Concord Avenue, Cambridge, MA 02138, USA. {\tt
eli@eecs.harvard.edu.} Research supported in part
by a Radcliffe Fellowship.}
\and Madhu Sudan%
\thanks{
MIT and Radcliffe IAS, The Stata Center Rm. G640, 32 Vassar Street,
Cambridge, MA 02139, USA.
{\tt madhu@mit.edu}. Research supported in part by
a Radcliffe Fellowship, and NSF Award
CCR-0219218.%
}
}
\maketitle

\fi

\newcommand{\poly}{\mathop{\rm poly}}
\newcommand{\Exp}{\mathop{\mathbf E}}

\newcommand{\TPC}{{\rm TPC}}
\newcommand{\tanner}{{\copyright}}
\newcommand{\csmall}{{C_{\rm small}}}
\newcommand{\eqdef}{\stackrel{\rm def}{=}}
\newcommand{\calp}{{\cal P}}

\begin{abstract}
We continue the investigation of locally testable codes, i.e.,
error-correcting codes for whom membership of a given word in the
code can be tested probabilistically by examining it
in very few locations. We give two general results on local
testability: First, motivated by the recently proposed notion of
{\em robust} probabilistically checkable proofs,
we introduce the notion of {\em robust} local
testability of codes. We relate this notion to a product of
codes introduced by Tanner,
and show a very simple composition lemma for this notion.
Next, we show that codes built by tensor products can
be tested robustly and somewhat locally, by applying a variant of
a test and proof technique introduced by Raz and Safra in the
context of testing low-degree multivariate polynomials (which are
a special case of tensor codes).

Combining these two results
gives us a generic construction of codes of
inverse polynomial rate, that are testable with
poly-logarithmically
many queries. We note these locally
testable tensor codes can be obtained
from {\em any} linear error correcting code with good distance.
Previous results on local
testability, albeit much stronger quantitatively, rely heavily on
algebraic properties of the underlying codes.
\end{abstract}

\section{Introduction}

Locally testable codes (LTCs) are error-correcting codes
that admit highly efficient probabilistic tests of membership.
Specifically, an LTC has a tester that makes a small number of
oracle accesses into an oracle representing a given word $w$,
accepts if $w$ is a codeword, and rejects with constant
probability if $w$ is far from every codeword.
LTCs are combinatorial counterparts of probabilistically checkable
proofs (PCPs), and were defined in \cite{FriedlS,RubinfeldS,Arora:thesis},
and their study was revived in~\cite{GS}.

Constructions of locally testable  codes typically come in two
stages. The first stage is algebraic and gives local tests for
algebraic codes, usually based on multivariate polynomials. This
is based on a rich collection of results on ``linearity
testing''
or ``low-degree
testing''
~\cite{AKKLR,ALMSS,AroraSafra,AroraSudan,BFLS,BFL,BCHKS,BGLR,BSVW,BLR,FGLSS,FHS,FriedlS,GS,PS,RubinfeldS}.
This first stage
either yielded codes of poor rate (mapping
$k$ information symbols to codewords of length $\exp(k)$)
as in~\cite{BLR},
or yielded codes over large alphabets as in~\cite{RubinfeldS}.
To reduce the alphabet size, a second stage of ``composition'' is
then applied. In particular, this is done in~\cite{GS,BSVW,BGHSV}
to get code mapping $k$ information bits to codewords of length
$k^{1+o(1)}$, over the {\em binary alphabet}. This composition
follows the lines of PCP composition introduced in \cite{AroraSafra},
but turns out to be fairly complicated, and in most cases, even more
intricate than PCP composition. The one exception is in \cite[Section 3]{GS},
where the composition is simple, but based
on very specific properties of the codes used.
Thus while the resulting constructions are surprisingly strong, the
proof techniques are somewhat complex.

In this paper, we search
for simple and general results related to local testing. A generic
(non-algebraic) analysis of low-degree tests appears in \cite{GSafra}, and
a similar approach to PCPs appears in \cite{DR}.
Specifically, we search for generic (non-algebraic) ways of
getting codes, possibly over large alphabets, that can be tested
by relatively local tests, as a substitute for algebraic ways.
And we look for simpler composition lemmas. We make some
progress in both directions. We show that the ``tensor product''
operation, a classical operation that takes two codes and produces
a new one, when applied to linear codes gives codes that are
somewhat locally testable (See Theorem~\ref{thm:m-product}).
To simplify the second stage, we strengthen the notion of
local testability to a ``robust'' one. This step is motivated
by an analogous step taken for PCPs in \cite{BGHSV}, but is naturally
formulated in our case using the ``Tanner Product'' for
codes~\cite{tanner}.
Roughly speaking, a ``big" Tanner Product code of block-length $n$ is
defined by a ``small" code
of block-length $n'=o(n)$ and a collection of subsets $S_1,\ldots,S_m
\subset [n]$, each of
size $n'$. A word is in the big code if and only if its projection to
every subset $S_i$ is a word of the small code. Tanner Product codes
have a natural local test associated with them: to test if a word $w$
is a codeword of the big code, pick a random subset $S_j$ and
verify that $w$ restricted to $S_j$ is a codeword of the small code.
The normal soundness condition would expect that if $w$ is far
from every codeword, then for a constant fraction of such restrictions,
$w$ restricted to $S_j$ is not a codeword of the small code.
Now the notion of robust soundness strengthens this condition
further by expecting that if $w$ is far from every codeword, then
many (or most) projections actually lead to words that are {\em far}
from codewords of the small code.
In other words, a code is robust if
global distance
(from the large code) translates into (average) local distance (from the
small code). A simple, yet
crucial observation is that robust codes compose naturally. Namely, if the
small code is itself locally testable by a robust test
(with respect to a tiny code, of block-length $o(n')$),
then distance from the large code (of block-length $n$)
translates to distance from the tiny code, thus reducing query complexity
while maintaining
soundness.
By viewing a tensor product as a robust Tanner product code,
we show that a $(\log N/\log \log N)$-wise tensor
product of {\em any} linear code of length $n = \poly \log N$ and
relative distance $1 - \frac{1}{\log N} = 1 -
\frac{1}{n^{\epsilon}}$, which yields a code of length $N$ and
polynomial rate, is testable with $\poly(\log N)$ queries
(Theorem~\ref{thm:final}). Once again, while stronger theorems
than the above have been known since \cite{BFLS}, the generic
nature of the result above might shed further light on the notion
of local testability.

\paragraph{Organization.}
We give formal definitions and mention our main theorems in
Section~\ref{sec:define}.
In Section~\ref{sec:tensor} we analyze the basic tester for
tensor product codes.
Finally in Section~\ref{sec:tanner} we describe our composition
and analyze some tests based on our composition lemma.

\section{Definitions and Main Results}
\label{sec:define}

Throughout this paper $\Sigma$ will denote a finite
alphabet, and in fact a finite field.
For positive integer $n$, let $[n]$ denote the set
$\{1,\ldots,n\}$. For a sequence $x \in \Sigma^n$
and $i \in [n]$,
we will let $x_i$ denote the $i$th element of the sequence.
The Hamming distance between strings $x,y \in \Sigma^n$,
denoted $\Delta(x,y)$, is the number of $i \in [n]$
such that $x_i \ne y_i$.
The relative distance between $x,y \in \Sigma^n$,
denoted $\delta(x,y)$, is the ratio $\Delta(x,y)/n$.

A code $C$ of length $n$ over $\Sigma$ is a subset of $\Sigma^n$.
Elements of $C$ are referred to as codewords.
When $\Sigma$ is a field, one may think of $\Sigma^n$ as a
vector space.
If $C$ is a linear subspace of
the vector space $\Sigma^n$, then $C$ is called a linear code.
The crucial parameters of a code, in addition to its length
and the alphabet, are its dimension (or information length)
and its distance, given by $\Delta(C) = \min_{x \ne y \in C}
\{\Delta(x,y)\}$.
A linear code of dimension $k$, length $n$, distance $d$
over the alphabet $\Sigma$ is denoted an $[n,k,d]_{\Sigma}$
code.
For a word $r \in \Sigma^n$ and a code $C$, we let
$\delta_C(r) = \min_{x \in C} \{\delta(r,x)\}$.
We say $r$ is
$\delta'$-proximate to $C$ ($\delta'$-far from $C$, respectively)
if $\delta_C(r)\geq \delta'$ ($\delta_C(r)\geq \delta'$,
respectively).

Throughout this paper, we will be working with infinite families
of codes, where their performance will be measured as a function
of their length.

\begin{definition}[Tester]
A tester $T$ with query complexity $q(\cdot)$ is a probabilistic
oracle machine that when given oracle access to a string
$r \in \Sigma^n$, makes $q(n)$ queries to the oracle for
$r$ and returns an accept/reject verdict.
We say that $T$ tests a code $C$ if
whenever $r \in C$, $T$ accepts with probability one;
and when $r \not\in C$, the tester rejects with
probability at least $\delta_C(r)/2$.
A code $C$ is said to be locally testable with $q(n)$ queries if there
is a tester for $C$ with query complexity $q(n)$.
\end{definition}

When referring to oracles representing vectors in $\Sigma^n$,
we emphasize the queries by denoting the response of the $i$th
query by $r[i]$, as opposed to $r_i$.
Through this paper we consider only non-adaptive testers, i.e.,
testers that use their internal randomness $R$ to generate $q$ queries
$i_1,\ldots,i_q \in [n]$ and a predicate $P:\Sigma^q \to \{0,1\}$
and accept iff $P(r[i_1],\ldots,r[i_q]) = 1$.

Our next definition is based on the notion of Robust PCP verifiers
introduced by \cite{BGHSV}. We need some terminology first.

Note that a tester $T$ has two inputs: an oracle for a received
vector $r$, and a random string $s$. On input the string $s$
the tester generates queries $i_1,\ldots,i_q \in [n]$ and
fixes circuit $C = C_s$ and accepts if $C(r[i_1],\ldots,r[i_q]) = 1$.
For oracle $r$ and random string $s$, define the robustness
of the tester $T$ on $r,s$, denoted
$\rho^{T}(r,s)$, to be the minimum, over strings $x$ satisfying
$C(x) = 1$, of relative distance of
$\langle r[i_1],\ldots,r[i_q]\rangle$ from $x$.
We refer to the quantity $\rho^{T}(r) \eqdef \Exp_s[\rho^{T}(r,s)]$
as the expected robustness of $T$ on $r$.
When $T$ is clear from context, we skip the superscript.

\begin{definition}
[Robust Tester]
A tester $T$ is said to be $\alpha$-robust for
a code $C$ if for every $r \in C$, the tester accepts w.p. one,
and
for every $r \in \Sigma^n$,
$\rho^{T}(r) \geq \alpha\cdot \delta_C(r)$.
\end{definition}

Having a robust tester for a code $C$
implies the existence of a tester for $C$, as illustrated
by the following proposition. 
\ifnum\full=0
It's proof appears in the full version of the paper \cite{bss-product-eccc}.
\fi

\begin{proposition}
\label{prop:amplify}
If a code $C$ has a $\alpha$-robust tester $T$ for $C$ making $q$
queries, then
it is locally testable with $O(q/\alpha)$ queries.
\end{proposition}

\ifnum\full=1

\begin{proof}
Let $c = \lceil\alpha^{-1}\rceil$.
The local tester $T'$
for $C$ is obtained by invoking $T$ $c$ times and accepting
if all invocations accept.
Consider a word $r$ with $\delta_C(r) = \delta$.
For at least $\delta/c \leq \alpha\cdot\delta$ 
fraction of the choices of random
strings $s$ of $T$, it must be that $\rho^{T}(r,s) > 0$
and $T$ rejects.
Thus the probability that $T'$ does not reject in any of the
$c$ repetitions is at most
\begin{eqnarray*}
(1 - \delta/c)^{c}
& \leq & 1 - c (\delta/c) + \binom{c}2 (\delta/c)^2 ~~\mbox{(By
Inclusion-Exclusion)} \\
& \leq & 1 - \delta + c^{2}/2 (\delta/c)^2 \\
& = &  1 - \delta + \delta^2/2 \\
& \leq & 1 - \delta + \delta/2 \\
& = & 1 - \delta/2
\end{eqnarray*}
Thus words at distance $\delta$ from codewords are
rejected with probability at least $\delta/2$.
\end{proof}

The previous proposition shows that large 
robustness leads to small query complexity. However, there is a limit to 
the size of the robustness parameter as shown in the next claim.

\begin{proposition}\label{prop:c-bound}
If $T$ is a $\alpha$-robust tester for a linear code $C\subset \Sigma^n$ 
with minimal (non-relativized) distance at least two, then $\alpha\leq 1$.
\end{proposition}

\begin{proof}
W.l.o.g. $T$ is a non-adaptive, i.e. the set of queries 
performed by $T$ does not depend on the received word $r$ (only on the 
randomness $s$) \cite{BHR}. Let $T_1,\ldots , T_S$ be the set of possible
tests performed by $T$, let $p_j$ be the probability $T_j$ is performed, and
let $q_j$ be the query complexity of $T_j$. Let $S_i$ be the set of tests
that query $i\in [n]$ and let ${\rm wt}(i)=\sum_{j\in S_i} p_j /q_j$ be the
weight of $i\in [n]$. There must be some $i$ with weight $\leq 1/n$ because
the sum of weights is one. Look at the word $r$ that is zero everywhere
but on the $i^{\rm th}$ coordinate, where it is one. On the one hand
$\delta_{C}(r)=1/n$, because $C$ is a linear code of minimal distance $>1$.
On the other hand, the robustness of $\rho^T(r)={\rm {wt}}(i)\leq 1/n$.
Thus, the robustness parameter is at most one.
\end{proof}

\fi

The main results of this paper focus on robust local testability
of certain codes. For the first result, we need to describe
the tensor product of codes.

\paragraph{Tensor Products and Local Tests}

Recall that an $[n,k,d]_\Sigma$ linear code $C$ may
be represented
by a $k \times n$ matrix $M$ over $\Sigma$ (so that
$C = \{x M | x \in \Sigma^k\}$). Such a matrix $M$ is called a
generator of $C$.
Given an $[n_1,k_1,d_1]_\Sigma$ code $C_1$ with generator $M_1$
and an $[n_2,k_2,d_2]_\Sigma$ code $C_2$ with generator $M_2$,
their tensor product (cf. \cite{MS},
\cite[Lecture 6, Section 2.4]{Sudan:lecture}), denoted
$C_1 \otimes C_2 \subseteq \Sigma^{n_2\times n_1}$, is the code
whose codewords may be viewed as $n_2\times n_1$ matrices given
explicitly by the set
$\{M_2^T X M_1 | X \in \Sigma^{k_2 \times k_1}\}$.
It is well-known that
$C_1 \otimes C_2$ is an
$[n_1n_2, k_1k_2, d_1d_2]_{\Sigma}$ code.

Tensor product codes are interesting to us in that they are a
generic construction of codes with ``non-trivially'' local
redundancy. To elaborate, every linear code of dimension $k$
does have redundancies of size $O(k)$, i.e., there exist
subsets of $t = O(k)$ coordinates where the code does not
take all possible $\Sigma^t$ possible values. But such redundancies
are not useful for constructing local tests; and unfortunately
generic codes of length $n$ and dimension $k$ may not have any
redundancies of length $o(k)$. However, tensor product codes
are different in that the tensor product of an $[n,k,d]_{\Sigma}$
code $C$ with itself leads to a code of dimension $k^2$ which is
much larger than the size of redundancies which are $O(k)$-long,
as asserted by the following proposition.

\begin{proposition}
\label{prop-one}
A matrix $r \in \Sigma^{n_2 \times n_1}$ is a codeword of
$C_1 \otimes C_2$ if and only if
every row is a codeword of $C_1$ and every column
is a codeword of $C_2$.
\end{proposition}

In addition to being non-trivially local, the constraints enumerated
above are also redundant, in that it suffices to insist that
all columns are codewords of $C_2$ and only $k_2$ (prespecified)
rows are codewords of $C_1$. Thus the insistence that other rows
ought to be codewords of $C_1$ is redundant, and leads to the hope
that the tests may be somewhat robust. Indeed we may hope that the
following might be a robust test for $C_1 \otimes C_2$.

\begin{quote}
{\bf Product Tester:} Pick $b \in \{1,2\}$ at random and $i \in [n_b]$
at random. Verify that $r$ with $b$th coordinate restricted to $i$
is a codeword of $C_{3-b}$.
\end{quote}

While it is possible to show that the above is a reasonable tester for
$C_1 \otimes C_2$, it remains open if the above is a robust tester
for $C_1 \otimes C_2$. (Note that the query complexity of the test
is $\max\{n_1,n_2\}$, which is quite high. However if the test
were robust, there would be ways of reducing this query complexity
in many cases, as we will see later.)

Instead, we consider higher products of codes,
and give a tester based on an idea from the work of
Raz and Safra~\cite{RazSaf}. Specifically, we let $C^m$ denote the
code $\underbrace{C \otimes \cdots \otimes C}_m$. We consider
the following test for this code:

\begin{quote}
{\bf $m$-Product Tester:} Pick $b \in [m]$ and $i \in [n]$
independently and uniformly at random.
Verify that $r$ with $b$th coordinate restricted to $i$
is a codeword of $C^{m-1}$.
\end{quote}

Note that this tester makes $N^{1 - \frac1m}$ queries to test
a code of length $N = n^m$. So its query complexity gets worse
as $m$ increases. However, we are only interested in the performance
of the test for small $m$ (specifically $m = 3,4$).
We show that the test is a robust tester for $C^m$ for every $m\geq 3$.
Specifically, we show

\begin{theorem}
\label{thm:m-product}
For a positive integer $m$ and $[n,k,d]_{\Sigma}$-code $C$,
such that $\left(\frac{d-1}n\right)^m \geq \frac78$,
$m$-Product Tester is $2^{-16}$-robust for $C^m$.
\end{theorem}

This theorem is proven in Section~\ref{sec:tensor}.
Note that the robustness is a constant, and the theorem only
needs the fractional distance of $C$ to be sufficiently large
as a function of $m$. In particular a fractional distance of
$1 - \frac{1}{O(m)}$ suffices. Note that such a restriction
is needed even to get the fractional distance of $C^m$ to be
constant.

The tester however makes a lot of queries,
and this might seem to make this result uninteresting
(and indeed one doesn't have to work so hard to get a
non-robust tester with such query complexity). However, as we note
next, the query complexity of robust testers can be reduced
significantly under some circumstances. To describe this we
need to revisit a construction of codes introduced by
Tanner~\cite{tanner}.

\paragraph{Tanner Products and Robust Testing}
\ifnum\full=1
The robustness of the $m$-Product Tester above seems to be
naturally related to the fact that the tester's predicates
are testing if the queried points themselves belong to a smaller
code. (In the case of the $m$-Product Tester, it verifies that
the symbols it reads give a codeword of the code $C^{m-1}$.)
The notion that a bigger code (such as $C^m$) may be specified
by requiring that certain projections of a word fall in a smaller
code (such as $C^{m-1}$) is not a novel one. Indeed this idea
goes back to the work of Tanner~\cite{tanner}, who defined this
notion in its full generality and considered big codes obtained
by a ``product'' of a bipartite graph with a small code.
This notion is commonly referred
to in the literature as the Tanner Product, and we define it next.
\fi

For integers $(n,m,t)$ an $(n,m,t)$-ordered bipartite graph
is given by $n$ left vertices $[n]$, and $m$ right vertices, where
each right vertex has degree $t$ and the neighborhood of a right
vertex $j \in [m]$ is ordered and given by a sequence $
\ell_j = \langle\ell_{j,1},\ldots,\ell_{j,t}\rangle$ with
$\ell_{j,i} \in [n]$.

A Tanner Product Code (TPC), is specified by an $[n,m,t]$
ordered bipartite graph $G$ and a code $\csmall \subseteq \Sigma^t$.
The product code, denoted $\TPC(G = \{\ell_1,\ldots,\ell_m\},\csmall)
\subseteq \Sigma^n$, is the set
$$\{r \in \Sigma^n~|~
r|_{\ell_j} \eqdef \langle r_{\ell_{j,1}},\ldots,r_{\ell_{j,t}} \rangle
\in \csmall,~\forall j \in [m]\}.$$

Notice that the Tanner Product naturally suggests a test for a code.
``Pick a random right vertex $j \in [m]$ and verify that $r|_{\ell_j}
\in \csmall$.''
Associating this test with such a pair $(G,\csmall)$, we say that
the pair is $\alpha$-robust if the associated test
is a $\alpha$-robust tester for $\TPC(G,\csmall)$.

The importance of this representation of tests comes from the
composability of robust tests coming from Tanner Product Codes.
Suppose $(G,\csmall)$ is $\alpha$-robust
and $\csmall$ is itself a Tanner Product Code, $\TPC(G',\csmall')$
where $G'$ is an $(d,m',t')$-ordered bipartite graph and
$(G',\csmall')$ is $\alpha'$-robust.
Then $\TPC(G,\csmall)$ has an
$\alpha\cdot\alpha'$-robust
tester that makes only $t'$ queries. (This fact is completely
straightforward and proven in Lemma~\ref{lem:comp}.)

This composition is especially useful in the context of
tensor product codes. For instance, the tester for $C^4$ is
of the form $(G,C^3)$, while $C^3$ has a robust tester of the
form $(G',C^2)$. Putting them together gives a tester for
$C^4$, where the tests verify appropriate projections are
codewords of $C^2$. The test itself is not surprising, however
the ease with which the analysis follows is nice. (See
Lemma~\ref{lem:fourtwo}.) Now the generality of the tensor
product tester comes in handy as we let $C$ itself be $C'^2$
to see that we are now testing $C'^8$ where tests verify some
projections are codewords of $C'^4$. Again composition allows
us to reduce this to a $C'^2$-test. Carrying on this way we
see that we can test any code of the form $C^{2^t}$ by verifying
certain projections are codewords of $C^2$. This leads to a simple
proof of the following theorem about the testability of
tensor product codes.

\begin{theorem}
\label{thm:final}
Let $\{C_i\}_i$ be any infinite family of codes with
$C_i$ a $[n_i,k_i,d_i]_{\Sigma_i}$ code,
with $n_i = p(k_i)$ for some polynomial $p(\cdot)$.
Further, let $t_i$ be a sequence of integers such
that $m_i = 2^{t_i}$ satisfies $d_i/n_i \geq 1 - \frac{1}{7m_i}$.
Then the sequence of codes $\{C'_i = C_i^{m_i}\}_i$ is a sequence
of codes of inverse polynomial rate and constant relative distance
that is locally testable with polylogarithmic number of queries.
\end{theorem}

This theorem is proven in Section~\ref{sec:tanner}.
We remark that it is possible to get code families $C_i$ such as
above using Reed-Solomon codes, as well as algebraic-geometric
codes.

\section{Testing Tensor Product Codes}
\label{sec:tensor}

Recall that in this section we wish to prove
Theorem~\ref{thm:m-product}. We first reformulate this
theorem in the language of Tanner products.

Let $G^n_m$ denote the graph that corresponds to the tests of
$C^m$ by the $m$-Product Tester, where $C \subseteq \Sigma^n$.
Namely $G^n_m$ has $n^m$ left vertices labelled by elements of
$[n]^m$. It has $m \cdot n$ right vertices labelled $(b,i)$ with
$b \in [m]$ and $i \in [n]$. Vertex $(b,i)$ is adjacent to all
vertices $(i_1,\ldots,i_m)$ such that $i_b = i$.
The statement of Theorem~\ref{thm:m-product} is equivalent to
the statement that $(G^n_m,C^{m-1})$ is $2^{-16}$-robust,
provided $\left(\frac{d-1}n\right)^m \geq \frac78$.
The completeness of the theorem follows
from Proposition~\ref{prop-one}, which implies
$C^m = \TPC(G^n_m,C^{m-1})$.
For the soundness, we first introduce some notation.

Consider the code $C_1\otimes\cdots\otimes C_m$, where
$C_i = [n_i,k_i,d_i]_\Sigma$ code.
Notice that codewords of this code lie in
$\Sigma^{n_1\times \cdots \times n_m}$. The
coordinates of strings in
$\Sigma^{n_1\times \cdots \times n_m}$
are themselves $m$-dimensional vectors over the integers
(from $[n_1]\times \cdots \times [n_m]$).
For $r \in
\Sigma^{n_1\times \cdots \times n_m}$ and $i_1,\ldots,i_m$
with $i_j \in [n_j]$,
let $r[i_1,\ldots,i_m]$ denote
the $\langle i_1,\ldots,i_m \rangle$-th coordinate of $r$.
For $b \in [m]$, and $i \in [n_b]$, let
$r_{b,i} \in \Sigma^{n_1 \times \cdots \times n_{b-1} \times n_{b+1} \times \cdots \times n_m}$
be the vector obtained by
projecting $r$ to coordinates whose $b$th coordinate is $i$,
i.e.,
$r_{b,i} [i_1,\ldots,i_{m-1}]
= r[i_1,\ldots,i_{b-1},i,i_b,\ldots,i_{m-1}]$.

The following simple property about tensor product codes will
be needed in our proof.
\ifnum\full=0
It's proof appears in the full version of the paper \cite{bss-product-eccc}.
\fi

\begin{proposition}
\label{prop-two}
For $b \in \{1,\ldots,m\}$ let $C_b$ be an $[n_b,k_b,d_b]_\Sigma$ code,
and let $I_b$ be a set of cardinality at least $n_b - d_b + 1$.
Let $C'_b$ be the code obtained by the projection of $C_b$ to $I_b$.
Then every codeword $c'$ of ${\cal{C}}'=C'_1 \otimes \cdots \otimes C'_m$
can be extended to a unique codeword $c$ of 
${\cal{C}}=C_1 \otimes \cdots \otimes C_m$.
\end{proposition}

\ifnum\full=1
\begin{proof}
The projection of $C_b$ to $C'_b$ is bijective. It is surjective
because it is a projection, and it is injective because $|I_b|> n_b-d_b$.
So, the projection of ${\cal{C}}$ to ${\cal{C}}'$ is a bijection, because
both codes are of dimension $\prod_{b=1}^m d_b$.
Thus, every word in ${\cal{C}}'$ has a unique preimage in ${\cal{C}}$. 
\end{proof}
\fi

Recall that the $m$-Product tester picks a random $b \in [m]$
and $i \in [n]$ and verifies that $r_{b,i} \in C^{m-1}$.
Let $\rho(r,(b,i))$ denote the expected distance of the view 
of this tester when accessing oracle $r$ 
on random string $(b,i)$.
Note that $\rho(r,(b,i)) =
\delta_{C^{m-1}}(r_{b,i})$.
Let $\rho(r) = \Exp_{b,i}[\delta_{C^{m-1}}(r_{b,i})]$.
We wish to show for every $r$ that $\rho(r) \geq 2^{-16} \cdot 
\delta_{C^m}(r)$ or equivalently $\delta_{C^m}(r) \leq 2^{16}
\cdot \rho(r)$.

We start by first getting a crude upper bound on the
proximity of $r$ to $C^m$ and then we use the crude bound
to get a tighter relationship. To get the crude bound,
we first partition the random strings into two classes:
those strings $(b,i)$ for which $\rho(r,(b,i))$ is large,
and those for which it is small.
More precisely,
for $r \in \Sigma^{n^m}$ and a threshold $\tau \in [0,1]$,
define the $\tau$-soundness-error of $r$ to be the probability
that $\delta_{C^{m-1}}(r_{b,i}) > \tau$, when $b \in [m]$ and
$i \in [n]$ are chosen uniformly and independently.
Note that the $\sqrt{\rho}$-soundness error of $r$ is
at most $\sqrt{\rho}$ for $\rho = \rho(r)$.
We start by showing that $r$ is
$O(\tau + \epsilon)$-close (and thus also $O(\sqrt{\rho})$-close)
to some codeword of $C^m$.

\begin{lemma}
\label{lem:tensor}
If the $\tau$-soundness-error of $r$ is $\epsilon$ for
$\tau + 2\epsilon \leq \frac{1}{12}\cdot\left(\frac{d-1}n\right)^m$,
then
$\delta_{C^m}(r)
\leq 16 \cdot \left(\frac{n}d\right)^{m-1}\cdot(\tau+\epsilon)$.
\end{lemma}

\begin{proof}
For every $i \in [n]$ and $b \in [m]$, fix $c_{b,i}$ to
be a closest codeword from $C^{m-1}$ to $r_{b,i}$.
We follow the proof outline of Raz \& Safra~\cite{RazSaf} which when adapted
to our context goes as follows:
(1) Given a vector $r$ and an assignment of codewords
$c_{b,i} \in C^{m-1}$,
we define an ``inconsistency'' graph $G$. (Note that this graph
is {\em not} the same as the graph $G^n_m$ that defines the test
being analysed. In particular $G$ is related to the word $r$ being
tested.) (2) We show that the
existence of a large independent set in this graph $G$ implies
the proximity of $r$ to a codeword of $C^m$ (i.e., $\delta_{C^m}(r)$
is small).
(3) We show that this
inconsistency graph is sparse if the $\tau$-soundness-error is small.
(4) We show that the distance of $C$ forces the graph to be special
in that every edge is incident to at least one vertex whose degree
is large.

\paragraph{Definition of $G$.}
The vertices of $G$ are indexed by pairs $(b,i)$ with $b \in [m]$
and $i \in [n]$. Vertex $(b_1,i_1)$ is adjacent to $(b_2,i_2)$ if
{\em at least} one of the following conditions hold:
\begin{enumerate}
\item $\delta_{C^{m-1}}(r_{b_1,i_1}) > \tau$.
\item $\delta_{C^{m-1}}(r_{b_2,i_2}) > \tau$.
\item $b_1 \ne b_2$ and $c_{b_1,i_1}$ and $c_{b_2,i_2}$ are
    inconsistent, i.e., there exists some element
    $j = \langle j_1,\ldots,j_m \rangle
    \in [n]^m$, with $j_{b_1} = i_1$ and $j_{b_2} = i_2$ such that
    $c_{b_1,i_1} [j^{(1)}] \ne c_{b_2,i_2}[j^{(2)}]$, where
    $j^{(c)} \in [n]^{m-1}$ is the vector $j$ with its $b_c$th coordinate
    deleted.
\end{enumerate}

\paragraph{Independent sets of $G$ and proximity of $r$.}
It is clear that $G$ has $mn$ vertices. We claim next that
if $G$ has an independent set $I$ of size at least
$m(n - d) + d + 1$
then $r$ has distance at most
$1 - (|I|/(mn))(1 - \tau)$ to $C^m$.

Consider an independent set $I = I_1 \cup \cdots \cup I_m$ in $G$
with $I_b$ of size $n_b$ being the set of vertices of the
form $(b, i), i\in [n]$. W.l.o.g. assume
$n_1 \geq \cdots \geq n_m$. Then, we have $n_1,n_2 > n-d$
(or else even if $n_1=n$ and $n_2=n-d$
we'd only have $\sum_b n_b \leq n + (m-1)(n-d)$).
We consider the partial vector $r'
\in \Sigma^{I_1 \times n \times \cdots \times n}$ defined as
$r'[i,j_2,\ldots,j_m] = c_{1,i}[j_2,\ldots,j_m]$ for $i \in I_1$,
and $j_2,\ldots,j_m \in [n]$.
We show that $r'$ can be extended into a codeword of $C^m$
and that the extended word is close to $r$ and this will
give the claim.

First, we show that any extension of $r'$ is close to $r$: This is
straightforward
since on each coordinate $i \in I_1$, we have $r$ agrees with
$r'$ on $1 - \tau$ fraction of the points. Furthermore
$I_1/n$ is at least $|I|/(mn)$ (since $n_1$ is the largest).
So we have that $r'$ is at most $1 - (|I|/(mn))(1 - \tau)$
far from $r$.

Now we prove that $r'$ can be extended into a codeword of $C^m$.
Let $C_b = C|_{I_b}$ be the projection (puncturing) of
$C$ to the coordinates in $I_b$.
Let $r''$ be the projection of $r'$ to the coordinates in
$I_1 \times I_2 \times [n] \times \cdots \times [n]$.
We will argue below that $r''$ is a codeword of $C_1 \otimes
C_2 \otimes C^{m-2}$, by considering its projection to axis-parallel
lines and claiming all such projections yield codewords of the
appropriate code.
Note first
that the restriction of $r'$ to any line parallel to the
$b$-th axis is a codeword of $C$, for every $b \in \{2,\ldots,m\}$,
since $r'_{1,i}$ is a codeword of $C^{m-1}$ for every $i \in I_1$.
Thus this continues to hold for $r''$ (except that now the projection
to a line parallel to the 2nd coordinate axis is a codeword of
$C_2$).
Finally, consider a line parallel to the first axis, given by
restricting the other coordinates to $\langle i_2,\ldots,i_m \rangle$,
with $i_2 \in I_2$. We claim that for every $i_1 \in I_1$,
$r''[i_1,\ldots,i_m] = c_{2,i_2}[i_1,\ldots,i_m]$. This follows from
the fact that the vertices $(1,i_1)$ and $(2,i_2)$ are not adjacent to
each other and thus implying that $c_{1,i_1}$ and $c_{2,i_2}$ are
consistent with each other.
We conclude that the restriction of $r''$ to every axis parallel line
is a codeword of the appropriate code, and thus
(by Proposition~\ref{prop-one}), $r''$ is a codeword of $C_1\otimes
C_2 \otimes C^{m-2}$.
Now applying Proposition~\ref{prop-two} to the code
$C_1\otimes C^{m-1}$ and its projection $C_1 \otimes C_2
\otimes C^{m-2}$ we get that there exists a unique extension of
$r''$ into a codeword $c'$ of the former.
We claim
this extension is exactly $r'$ since for every $i \in I_1$,
$c'_{1,i}[j,k] = r'[i,j,k]$.
Finally applying Proposition~\ref{prop-two} one more time,
this time to the code $C^m$ and its projection
$C_1 \otimes C^{m-1}$, we find that $r'=c'$ can be extended
into a codeword of the former. This concludes the proof of this
claim.

\paragraph{Density of $G$.}
We now see that the small $\tau$-soundness-error of the test
translates into a small density $\gamma$ of edges in $G$.
Below, we refer to pairs $(b,i)$ with $b \in [m]$ and $i \in [n]$
as ``planes'' (since they refer to $(m-1)$-dimensional planes in
$[n]^m$) and refer to elements of $[n]^m$ as ``points''. We
say a point $p = \langle p_1,\ldots,p_m \rangle$ lies on
a plane $(b,i)$ if $p_b = i$.
Now consider the following test: Pick two random planes
$(b_1,i_1)$ and $(b_2,i_2)$ subject to the constraint $b_1
\ne b_2$ and pick a random point $p$ in the intersection of
the two planes and verify that $c_{b_1,i_1}$ is consistent
with $r[p]$. Let $\kappa$ denote the rejection probability of
this test. We bound $\kappa$ from both sides.

On the one hand we have that the rejection probability is
at least the probability that we pick two planes that are
$\tau$-robust and incident to each other in $G$ (which is at least
$\frac{m\gamma}{m-1} - 2\epsilon$) and the probability that
we pick a point on the intersection at which the two
plane codewords disagree (at least $(d/n)^{m-2}$),
times the probability that the codeword that disagrees with
the point function is the
first one (which is at least $1/2$). Thus we get
$\kappa \geq \frac{d^{m-2}}{2(n)^{m-2}} \left(
\frac{m\gamma}{m-1} - 2\epsilon \right)$.

On the other hand we have that in order to reject it must
be the case that either $\delta_{C^{m-1}}(r_{b_1,i_1}) > \tau$ (which
happens with probability at most $\epsilon$) or
$\delta_{C^{m-1}}(r_{b_1,i_1}) \leq \tau$ and $p$ is
such that $r_{b_1,i_1}$ and $c_{b_1,i_1}$ disagree at
$p$ (which happens with probability at most $\tau$).
Thus we have
$\kappa \leq \tau + \epsilon$.
Putting the two together we have
$\gamma \leq \frac{m-1}{m}\left(
2\epsilon + \frac{2n^{m-2}}{d^{m-2}} (\tau + \epsilon)
\right)$.

\paragraph{Structure of $G$.}

Next we note that every edge of $G$ is incident to at
least one high-degree vertex. Consider a pair of planes
that are adjacent to each other in $G$. If either of the
vertices is not $\tau$-robust, then it is adjacent to every vertex
of $G$. So assume both are $\tau$-robust.

W.l.o.g., let these
be the vertices $(1,i)$ and $(2,j)$.
Thus the codewords $c_{1,i}$
and $c_{2,j}$ disagree on the $(m-2)$-dimensional surface
with the first two coordinates restricted to $i$ and $j$ respectively.
Now let
$S = \{\langle k_3,\ldots,k_m \rangle ~|~
c_{1,i}[j,k_3,\ldots,k_m] \ne c_{2,j}[i,k_3,\ldots,k_m]\}$ be the set
of disagreeing tuples on this line. By the distance of $C^{m-2}$ we
know $|S| \geq d^{m-2}$.
But now if we consider the vertex $(b,k_b)$
in $G$ for $b \in \{3,\ldots,m\}$ and $k_b$
such that there exists $k_1,\ldots,k_{m-2}$ satisfying
$k = (k_1,\ldots,k_{m-2}) \in S$, it must be adjacent at least one of
$(1,i)$ or $(2,j)$ (it can't agree with both at the point
$(i,j,k)$. Furthermore, there exists $d$ such $k_b$'s for every
$b \in \{3,\ldots,m\}$.
Thus the sum of the degrees of $(1,i)$ and $(2,j)$ is
at least $(m-2)d$, and so at least one has degree at least $(m-2)d/2$.

\paragraph{Putting it together.}
 From the last paragraph above, we have that the set of vertices of
degree less than $(m-2)d/2$ form an independent set in the graph $G$.
The fraction of vertices of degree at least $(m-2)d/2$ is at most
$2(\gamma m n)/((m-2)d)$.
Thus we get that if $mn\cdot \left(1 - 2(\gamma m n)/((m-2)d)\right)
\geq m(n-d) + d + 1$, then $r$ is $\delta$-proximate to $C^m$ for
$\delta \leq \tau + (1 - \tau)\cdot 2(\gamma m n)/((m-2)d)$.
The lemma now follows by simplifying the expressions above,
using the upper bound on $\gamma$ derived earlier. 
\ifnum\full=1
Details below.

We first focus on the condition $|I| \geq m(n-d)+d+1$. It suffices
to prove that
\begin{eqnarray*}
mn\cdot \left(1 - 2(\gamma m n)/((m-2)d)\right)
& \geq & m(n-d) + d + 1 \\
\Leftrightarrow
(m-1)d - 1 & \geq & 2 \frac{\gamma m^2 n^2}{(m-2)d} \\
\Leftarrow
(m-1)(d - 1) & \geq & 2 \frac{\gamma m^2 n^2}{(m-2)d} \\
\Leftarrow
(m-1)(d - 1) & \geq & 2 \frac{m^2 n^2}{(m-2)d} \cdot \frac{m-1}{m}
\cdot \left(2 \epsilon +
            2\left(\frac{n}d\right)^{m-2} \cdot (\tau + \epsilon)\right)\\
\Leftarrow
(d - 1) & \geq & 2 \frac{m n^2}{(m-2)d} \cdot
\left(2 \epsilon +
            2\left(\frac{n}d\right)^{m-2} \cdot (\tau + \epsilon)\right)\\
\Leftarrow
(d - 1) & \geq & 2 \frac{m n^2}{(m-2)(d-1)} \cdot
\left(2 \epsilon +
            2\left(\frac{n}{d-1}\right)^{m-2} \cdot (\tau + \epsilon)\right)\\
\Leftarrow
\left(2 \epsilon + 2\left(\frac{n}{d-1}\right)^{m-2} \cdot (\tau + \epsilon)\right)
& \leq & \frac{(m-2)(d-1)^2}{2m n^2} \\
\Leftarrow
\left(2\left(\frac{n}{d-1}\right)^{m-2} \cdot (\tau + 2\epsilon)\right)
& \leq & \frac{(m-2)(d-1)^2}{2m n^2} \\
\Leftarrow
(\tau + 2\epsilon)
& \leq & \frac{m-2}{2m} \cdot \left(\frac{d-1}{n}\right)^m \\
\Leftarrow
(\tau + 2\epsilon)
& \leq & \frac1{12} \cdot \left(\frac{d-1}{n}\right)^m \\
\end{eqnarray*}

The above shows that the condition assumed in the lemma statement
indeed is sufficient to establish a large independent set. Next
we simplify the proximity bound obtained.
We have
\begin{eqnarray*}
\delta & \leq & \tau + (1 - \tau)\cdot \frac{2\gamma m n}{(m-2)d} \\
& \leq & \tau + \frac{2\gamma m n}{(m-2)d} \\
& \leq & \tau + \frac{2 m n}{(m-2)d}
\cdot \frac{m-1}{m} \cdot \left(2 \epsilon +
2\left(\frac{n}d\right)^{m-2}\cdot(\tau+\epsilon) \right) \\
& \leq & \tau + \frac{2 m n}{(m-2)d}
\cdot \frac{m-1}{m} \cdot
2\left(\frac{n}d\right)^{m-2}\cdot(\tau+2\epsilon) \\
& \leq & \tau + \frac{4 (m-1)}{m-2}
\cdot \left(\frac{n}d\right)^{m-1}\cdot(\tau+2\epsilon) \\
& \leq & \frac{4 (m-1)}{m-2}
\cdot \left(\frac{n}d\right)^{m-1}\cdot(2\tau+2\epsilon) \\
& \leq & \frac{8 (m-1)}{m-2}
\cdot \left(\frac{n}d\right)^{m-1}\cdot(\tau+\epsilon) \\
& \leq & 16 \cdot \left(\frac{n}d\right)^{m-1}\cdot(\tau+\epsilon). \\
\end{eqnarray*}
\else
The calculations needed to derive the simple expressions can be
found in the full version of this paper~\cite{bss-product-eccc}.
\fi
\end{proof}

Next we improve the bound achieved on the proximity of
$r$
by looking at the structure of the graph $G^n_m$ (the
graph underlying the $m$-Product tester) and its
``expansion''. Such improvements are a part of the standard
toolkit in the analysis of low-degree tests based on axis
parallel lines (see e.g., \cite{BFL,BFLS,FGLSS,FHS} etc.)
\ifnum\full=0
We omit the proof of this part from this version --- it may be
found in the full version of this paper~\cite{bss-product-eccc}.
We state the resulting lemma and show how Theorem~\ref{thm:m-product}
follows from it.
\else
We follow the proof outline of \cite{FHS} which in turn uses
a proof technique of \cite{BS}.

First, some notation: Fix $n$ and $m$ and the graph $G^n_m$.
Let $L$ and $R$ denote the left and
right vertices of $G^n_m$. Let $d_L$ and $d_R$ denote
its left and right degrees. And let $E$ denote the edges of
$H^n_2$. Note $|L| = n^m$, $|R| =
m n$, $d_L = m$ and
$d_R = n^{m-1}$. In particular, $d_L \cdot |L| = d_R \cdot |R|$.
For a set $A \subseteq L \cup R$, let
$\Gamma(A) = \{(u,v) \in E ~|~ u \in A, v \not\in A\}$.
Using this notation, we have the following:

\begin{lemma}
\label{lem:expand}
Fix $n,m\geq 3$ and let $L$, $R$ denote the two
partitions of the vertices of $G^n_m$ and let $d_L,d_R$
denote the left and right degrees.
Let $S \subseteq L$ and $T \subseteq R$ be such that
$\frac{|S|}{|L|} \leq \frac14$.
Then $|\Gamma(S \cup T)| \geq
\frac{d_L}{8}\cdot|S| + \frac{d_R}{8} \cdot |T|$.
\end{lemma}

\begin{proof}
We start with a simple observation that also allows us
to bound the size of $T$. Suppose, $|T| \geq |R|/2$.
Then the number of edges leaving $T$ is at least
$d_R \cdot |T| \geq d_R \cdot (|R|/2)$.
On the other hand the number of edges entering $S$ is
at most $d_L \cdot |S| \leq d_L \cdot (|L|/4)$.
Thus in this case, we have
\begin{eqnarray*}
\Gamma(S\cup T) & \geq & d_R \cdot (|R|/2) - d_L \cdot (|L|/4) \\
& = & d_R \cdot (|R|/4) \\
& = & d_R \cdot (|R|/8) + d_L \cdot (|L|/8) \\
& \geq & d_R \cdot (|S|/8) + d_L \cdot (|T|/8).
\end{eqnarray*}

We are thus reduced to the case where $|S|/|L|,|T|/|R| \leq \frac12$.
Here, we follow the proof of Babai and Szegedy~\cite{BS}.
\ifnum\full=1
(See also \cite{Lovasz:random-walks}).
\fi
The crucial fact needed to apply their proof is that the graph
$G^n_m$ is edge-transitive, i.e., for every pair of edges $e_1,e_2$
in $G^n_m$, there is an automorphism of $G^n_m$ that maps
$e_1$ to $e_2$. This fact is used as follows:
Let $A$ denote the set of all automorphisms
of $G^n_m$. Then if we consider any fixed edge $e \in H^n_t$
and all its images under automorphisms $A$ as a multiset, then
every edge of $G^n_m$ appears exactly the same number of times.

Armed with this fact, the proof proceeds as follows:
For every pair $u \in L$ and $v \in R$ define a canonical
shortest path $P_{u,v}$. Note that this path has length at most
three. Note that an automorphism from $A$ maps a path in $H^n_t$
to a path in $H^n_t$. Now consider the multiset $\calp$
of all paths obtained by taking the paths $P_{u,v}$ for every
$u,v$, and their automorphisms for every automorphism in $A$.
The cardinality of $\calp$ is thus $|A|\cdot |L|\cdot |R|$.
The symmetry over the edges implies that every edge in $E$ has
exactly the same number, say $N$, of paths from $\calp$ passing
through them. Since each path has at most three edges, we have
$N \leq \frac{3\cdot|A|\cdot|R|}{d_L}
= \frac{3\cdot|A|\cdot|L|}{d_R}$, or equivalently
$\frac{|A|}N \geq \frac{d_L}{3\cdot |R|} = \frac{d_R}{3 \cdot |L|}$.

Now consider the set of paths $\calp' \subseteq \calp$
whose endpoints involve exactly one element of $S \cup T$.
We have the cardinality of $\calp'$ equals $|A| \cdot
(|S| \cdot |\overline{T}| + |\overline{S}| \cdot |T|)$
(where $\overline{S} = L - S$ and $\overline{T} = R - T$).
On the other hand, we have $|\calp'| \leq N \cdot |\Gamma(S \cup T)|$

Combining the two we have
\begin{eqnarray*}
|\Gamma(S \cup T)| & \geq & \frac{1}N \cdot |\calp'| \\
& \geq & \frac{|A|}N \cdot
(|S| \cdot |\overline{T}| + |\overline{S}| \cdot |T|) \\
& \geq & \frac{d_L}{3\cdot |R|}\cdot|S|\cdot
|\overline{T}| + \frac{d_R}{3 \cdot |L|} \cdot
|\overline{S}| \cdot |T| \\
& \geq & \frac{d_L}{6}\cdot|S| + \frac{d_R}{6} \cdot |T|.
\end{eqnarray*}
This proves the lemma.
\end{proof}
\fi

\begin{lemma}
\label{lem:self-improv}
Let $m$ be a positive integer and
$C$ be an $[n,k,d]_{\Sigma}$ code
with the property $d^{m-1}/n^{m-1} \geq \frac{7}{8}$.
If $r \in \Sigma^{n^m}$ and $c \in C^m$
satisfy $\delta(r,c)\leq \frac14$
then $\delta(r,c) \leq 8\rho(r)$.
\end{lemma}

\ifnum\full=1

\begin{proof}
Let $L,R$ denote the two partitions of the vertices of
$G^n_m$. Note that the right vertices of $G^n_m$ are
of the form $(b,i)$, with $b \in [m]$ and $i \in [n]$.
Let $r_{b,i}$ denote the projection of $r$ to the
neighborhood of the right vertex $(b,i)$, and let
$c_{b,i}$ denote the projection of $c$ to the same.
Let $c'_{b,i}$ denote the codeword of $C^{m-1}$
closest to $r_{b,i}$.
Call an edge $(u,(b,i))$ of $G^n_m$ {\em bad} if $r$ and
$c'_{b,i}$ disagree at $u$.
Note that the fraction of bad edges equals $\rho(r)$.

We now lower bound $\rho(r)$ in terms of $\delta(r,c)$. For this part
we use Lemma~\ref{lem:expand}.
Let $S \subseteq L$ be the set of vertices
$(i_1,\ldots,i_m)$ for which
$r[i_1,\ldots,i_m] \ne c[i_1,\ldots,i_m]$.
Note that by assumption $|S|/|L| = \delta(r,c) \leq \frac14$.
Let $T \subseteq R$ be the set of vertices $(b,i)$
for whom $c_{b,i} \ne c'_{b,i}$.
By Lemma~\ref{lem:expand} we have $|\Gamma(S \cup T)|
\geq
\frac{d_L}{8}\cdot|S| + \frac{d_R}{8} \cdot |T|$.
We now claim that most of these edges are {\em bad}.

Consider first an edge $(u,(b,i))$ in $G^n_m$ from
$S$ to $\overline{T}$.
On the one hand $c'_{b,i} = c_{b,i}$ and on the other $r[u] \ne c[u]$.
This leads to a disgreement between $r$ and $c'$ at $u$ and so such
an edge is bad.
Next, consider an edge $(u,(b,i))$ from $u \in \overline{S}$ to $T$.
We do have $r[u] = c[u]$ and $c'_{b,i} \ne c_{b,i}$, but this doesn't
imply
that $(u,(b,i))$ is bad, since $c_{b,i}$ and $c'_{b,i}$ need not
disagree at $u$.
Indeed for every $(b,i) \in T$, there may be up to
$n^{m-1} - d^{m-1}$ edges $(u,(b,i))$ for which $c'_{b,i}$ and
$r$ agree at
$u$, but remaining edges out of $(b,i)$ are bad.
Discounting for these edges, we see that all but at most $(n^{m-1}
- d^{m-1})
\cdot |T|$ edges from $T$ to $\overline{S}$ are bad.
Thus we get that the number of bad edges is at least
$\frac{d_L}{8}\cdot|S| + \frac{d_R}{8} \cdot |T| -
(n^{m-1} - d^{m-1})\cdot |T|$.
Using $d_R = n^{m-1}$ and $d^{m-1}/n^{m-1} \geq \frac{7}{8}$,
we get
$\frac{d_R}{8} \cdot |T| - (n^{m-1} - d^{m-1})\cdot |T| \geq 0$.
Thus we get that the fraction of bad edges $\beta$ is at least
$\frac{1}{8}\cdot(|S|/|L|) = \frac{\delta(r,c)}{8}$.
We conclude $\delta(r,c)\leq 8\cdot \rho(r)$.
\end{proof}

We are now ready to put the pieces together to prove
Theorem~\ref{thm:m-product}.
\fi


\begin{proof}[Theorem~\ref{thm:m-product}]
Let $\alpha = 2^{-14}\cdot\left(\frac{d-1}{n}\right)^{2m}$.
We will prove that the $m$-Product
Tester is $\alpha$-robust for $C^m$.
Note that $\alpha \geq 2^{-16}$ as required for the theorem,
and $\sqrt{\alpha} \leq
\min\{\frac1{36}\cdot\left(\frac{d-1}n\right)^m,
\frac{1}{128}\cdot\left(\frac{d}n\right)^{m-1}\}$
(as will be required below).

The completeness (that codewords of $C^m$ have expected
robustness zero) follows from Proposition~\ref{prop-one}.
For the soundness,
consider any vector $r\in \Sigma^{n^m}$ and let $\rho = \rho(r)$.
If $\rho > \alpha$, then there there is nothing to
prove since $\rho/\alpha > 1 \geq \delta_{C^m}(r)$.
So assume $\rho  \leq \alpha$.

Note that $r$ has $\sqrt{\rho}$-soundness-error at most $\sqrt{\rho}$.
Furthermore, by the assumption on $\rho$, we have
$3\sqrt{\rho} \leq
3\sqrt{\alpha} \leq \frac1{12}\cdot \left(\frac{d-1}n\right)^m$
and so, by Lemma~\ref{lem:tensor}, we have
$\delta_{C^m}(r) \leq 16 \cdot \left(\frac{n}d\right)^{m-1}
\cdot 2 \cdot \sqrt{\rho}$.
Now using $\sqrt{\rho} \leq \sqrt{\alpha} \leq \frac{1}{128} \cdot
\left(\frac{d}n\right)^{m-1}$, we get
$\delta_{C^m}(r) \leq \frac14$.
Let $v$ be a codeword of $C^m$ closest to $r$. We now
have $\delta(r,v) \leq \frac14$ and
$\left(\frac{d}n\right)^{m-1} \geq \frac78$, and
so, by Lemma~\ref{lem:self-improv}, we get
$\delta_{C^m}(r) = \delta(r,v) \leq 8 \rho$.
This concludes the proof.
\end{proof}

\section{Tanner Product Codes and Composition}
\label{sec:tanner}

In this section we define the composition of two
Tanner Product Codes, and show how they preserve robustness.
We then use this composition to show how to test $C^m$
using projections to $C^2$.
\ifnum\full=0
All proofs of Lemmas in this Section appear in the
full version of the paper \cite{bss-product-eccc}.
\fi

\ifnum\full=1
\subsection{Composition}
\label{ssec:comp}
\fi

Recall that a Tanner Product Code is given by a pair $(G,\csmall)$.
We start by defining a composition of graphs that corresponds to
the composition of codes.

Given an $(N,M,D)$-ordered graph $G = \{\ell_1,\ldots,\ell_M\}$
and an additional $(D,m,d)$-ordered graph $G' = \{\ell'_1,\ldots,\ell'_m\}$,
their Tanner Composition, denoted $G \tanner G'$, is an
$(N,M\cdot m,d)$-ordered graph with adjacency lists
$\{\ell''_{j,j'} | j \in [M], j' \in [m]\}$, where
$\ell''_{(j,j'),i} = \ell_{j,\ell'_{j',i}}$.

\begin{lemma}[Composition]
\label{lem:comp}
Let $G_1$ be an $(N,M,D)$-ordered graph, and
$C_1 \subseteq \Sigma^D$ be a linear code with
$C = \TPC(G_1,C_1)$.
Further, let $G_2$ be an $(D,m,d)$-ordered graph
and $C_2 \subseteq \Sigma^d$ be a linear code such that
$C_1 = \TPC(G_2,C_2)$.
Then $C = \TPC(G_1 \tanner G_2, C_2)$ (giving a
$d$-query local test for $C$).
Furthermore if
$(G_1,C_1)$ is $c_1$-robust
and
$(G_2,C_2)$ is $c_2$-robust, then
$(G_1 \tanner G_2,C_2)$ is
$c_1 \cdot c_2$-robust.
\end{lemma}

\ifnum\full=1
\begin{proof}
We focus on the robustness of the $C$, as all other claims follow
immediately from definitions. Assume $w\in \Sigma^N$ has distance
$\delta$ from $C$. Then, since $C=\TPC(G_1,C_1)$ is $c_1$-robust,
the expected distance of a random "medium"-size test (of query size $D$)
is at least $\delta c_1$, so by the $c_2$-robustness of $C_1=\TPC(G_2,C_2)$ the expected
distance of the "small"-size test (of query complexity $d$) is at least
$\delta c_1\cdot c_2$ as claimed.

\end{proof}

\fi

\ifnum\full=1
\subsection{Testing a 4-Wise Tensor Product Code}
\fi

We continue by recasting the results of Section~\ref{sec:tensor}
in terms of robustness of associated Tanner Products.
Recall that $G^n_m$ denotes the graph that corresponds to
the tests of $C^m$ by the $m$-Product Tester,
where $C \subseteq \Sigma^n$.

Note that $G^n_m$ can be composed with $G^n_{m-1}$ and so on.
For $m' < m$, define $G^n_{m,m'} = G^n_m$ if $m' = m-1$ and
define $G^n_{m,m'} = G^n_m \tanner G^n_{m-1,m'}$ otherwise.
Thus we have that $C^m = \TPC(G^n_{m,m'},C^{m'})$.
The following lemma (which follows easily from
Theorem~\ref{thm:m-product} and Lemma~\ref{lem:comp}
gives the robustness of
$(G^n_{4,2},C^2)$.

\begin{lemma}
\label{lem:fourtwo}
Let $C$ be an $[n,k,d]_{\Sigma}$ code with
$(d-1/n)^4 \leq \frac78$.
Then $(G^n_{4,2},C^2)$ is $2^{-32}$-robust.
\end{lemma}

\ifnum\full=1
\begin{proof}
Since we have $(d-1/n)^4 \geq \frac{7}8$ we may
apply Theorem~\ref{thm:m-product} with $m = 3,4$
to get that $(G^n_4,C^3)$ and $(G^n_3,C^2)$ are
both $2^{-16}$-robust.
Since $C^3 = \TPC(G^n_3,C^2)$, we may apply
Lemma~\ref{lem:comp} to conclude that
$(G^n_{4,2} = G^n_{4} \tanner G^n_3, C^2)$ is
$2^{-32}$-robust.
\end{proof}

\subsection{Testing Tensor Products with $C^2$ tests}
\fi

Finally we define graphs $H^n_t$ so that
$C^{2^t} = \TPC(H^n_t,C^2)$. This is easily done recursively
by letting
$H^n_2 = G^n_{4,2}$ and letting
$H^n_t = G^{n^{2^{t-2}}}_{4,2} \tanner H^n_{t-1}$ for $t > 2$.
We now analyze the robustness of $(H^n_t,C^2)$.

\begin{lemma}
\label{lem:mtwo}
There exists a constant $\alpha >0 $ such that the following holds:
Let $t$ be an integer and $C$ be an $[n,k,d]_\Sigma$ code
such that $d - 1 \geq (1 - \frac{1}{10m})\cdot n$, for $m = 2^t$.
Then $(H^n_{t},C^2)$ is
$\alpha^t$-robust.
\end{lemma}

\ifnum\full=1
\begin{proof}
Note that the condition in the lemma implies $((d-1)/n)^m
\geq (1 - \frac{1}{5m})^m \geq e^{-0.1} \geq \frac78$.
This is the form in which
we use the condition.

We prove the lemma, for $\alpha = 2^{-32}$, by induction.
For the base case, we have
$(H^n_2 = G^n_{4,2},C^2)$ is $2^{-32}$-robust, by
Lemma~\ref{lem:fourtwo}. (Here we use the fact that
$(d-1/n)^4 \geq \frac78$ as needed.)

For the induction,
let $m = 2^t$.
and let $C' = C^{m/4}$.
Let $G_1 = G^{n^{m/4}}$, $C_1 = (C')^2$,
$G_2 = H^n_{t-1}$ and $C_2 = C^2$
Note that $H^n_t = G_1 \tanner G_2$ and $C_1 = \TPC(G_2,C_2)$.
Thus we can
bound the robustness of $(H^n_t,C_2)$
by bounding the robustness of $(G_1,C_1)$ and
$(G_2,C_2)$ and then
using Lemma~\ref{lem:comp}.
Note that $C_1 = (C')^2$ and $C'$ is a
$[n^{m/4},k^{m/4},d^{m/4}]_{\Sigma}$
code, where
$$\left(\frac{d^{m/4}-1}{n^{m/4}}\right)^4 \geq
\left(\frac{d-1}{n}\right)^m \geq \frac78.$$
Thus we can apply Lemma~\ref{lem:fourtwo} to
conclude $(G_1,C_1) = (G^n_{4,2},(C')^2)$ is
$\alpha$-robust for $\alpha=2^{-32}$.
By induction, we also have $(G_2,C_2) = (H^n_{t-1},C^2)$
is $\alpha^{t-1}$-robust.
By Lemma~\ref{lem:comp}, $(G_1\tanner G_2,C_2)$ is
$\alpha^t$-robust.
\end{proof}
\fi

We are ready to prove Theorem~\ref{thm:final}.

\begin{proof}[Theorem~\ref{thm:final}]
Let $\alpha$ be the constant given by Lemma~\ref{lem:mtwo}.
Fix $i$ and let $C = C_i$, $n = n_i$ etc. (i.e., we suppress the
subscript $i$ below).
Then $C^m$ is an $[N,K,D]_q$ code,
for $N = n^m$, $K = k^m$ and $D = d^m$.
Since $d/n \geq 1 - \frac{1}{2m}$, we have $C^m$
has relative distance $d^m/n^m \geq \frac12$.
Furthermore, the rate of the code is
inverse polynomial, i.e., $N = n^m = (p(k))^m \leq \poly(k^m) = \poly(K)$.
Finally, we have $C^m = \TPC(H^n_{\log_2 m},C^2)$,
where $(H^n_{\log_2 m}, C^2)$ is an
$\alpha^{\log_2 m}$-robust tester for
$C^m$
and this tester has query complexity $O(n^2)$.
From Proposition~\ref{prop:amplify} we get that there is a
tester for $C$ that makes $O(n^2/ \alpha^{O(\log_2 m)}) =
\poly \log N$ queries.
\end{proof}

\section*{Acknowledgments}

We wish to thank Irit Dinur, Oded Goldreich and Prahladh Harsha
for valuable discussions.

\bibliographystyle{plain}

\end{document}